\documentclass[letterpaper, 10 pt, conference]{ieeeconf}  

\IEEEoverridecommandlockouts                              

\usepackage{graphicx} 
\usepackage{booktabs}  
\usepackage{threeparttable}  

\title{\LARGE \bf
Brain Tumors Classification for MR images based on \\
Attention Guided Deep Learning Model
}

\author{Yuhao Zhang$^{1}$, Shuhang Wang$^{2}$, Haoxiang Wu$^{3}$, Kejia Hu$^{4,5,*}$, Shufan Ji$^{6,*}$  
\thanks{$^{1}$School of Sino-French Engineer, Beihang University, Beijing, 100191, China}
\thanks{$^{2}$Department of Radiology, Massachusetts General Hospital, Harvard Medical School, Boston, MA, 02138, USA}
\thanks{$^{3}$School of Medicine, Shanghai Jiao Tong University, 200025, China.}
\thanks{$^{4}$Center for Functional Neurosurgery, Ruijin Hospital, Shanghai Jiao Tong University School of Medicine, Shanghai, 200025, China.}
\thanks{$^{5}$Department of Neurosurgery, Ruijin Hospital, Shanghai Jiao Tong University School of Medicine, Shanghai, 200025, China.}
\thanks{$^{6}$School of Computer Science and Engineering, Beihang University, Beijing, 100191, China}
\thanks{$^{*}$Co-corresponding author; E-mail address: dockejiahu@gmail.com, jishufan@buaa.edu.cn}
}

\begin{document}

\maketitle
\thispagestyle{empty}
\pagestyle{empty}

\begin{abstract}
 In the clinical diagnosis and treatment of brain tumors, manual image reading consumes a lot of energy and time. In recent years, the automatic tumor classification technology based on deep learning has entered people's field of vision. Brain tumors can be divided into primary and secondary intracranial tumors according to their source. However to our best knowledge, most existing research on brain tumors are limited to primary intracranial tumor images and cannot classify the source of the tumor. In order to solve the task of tumor source type classification, we analyze the existing technology and propose an attention guided deep convolution neural network (CNN) model. Meanwhile, the method proposed in this paper also effectively improves the accuracy of classifying the presence or absence of tumor. For the brain MR dataset, our method can achieve the average accuracy of 99.18\% under ten-fold cross-validation for identifying the presence or absence of tumor, and 83.38\% for classifying the source of tumor. Experimental results show that our method is consistent with the method of medical experts. It can assist doctors in achieving efficient clinical diagnosis of brain tumors.
\end{abstract}

\section{INTRODUCTION}

According to the World Cancer Report provided by the World Health Organization (WHO), there are approximately 297 000 new cases of brain tumor cases worldwide in 2018. Brain tumors can be divided into two categories according to the origin: primary and secondary intracranial tumors. Primary intracranial tumors are tumors of origin that grow in the brain. And the types that generally occur based on the affected area are meningioma, glioma, and pituitary. Secondary intracranial brain tumors are tumors growing within the brain that has arisen from the spread of a malignant tumors elsewhere in the body. And common metastatic sites are colorectal, lung, and breast. Usually, the doctor will make a preliminary assessment of the situation through medical imaging technology. In recent years, magnetic resonance imaging (MRI) technology has been widely used to generate high-resolution images of brain structures. It has the advantages of diverse imaging methods and rich information. However, for a large number of MR images, it takes a long time to read. For this problem, automatic brain tumor classification is a possible solution.

CNNs based on deep learning are currently the best way to classify data in the field of computer vision. In recent years, CNNs have been widely used for medical image diagnosis. Unlike traditional methods, CNNs preserve the input’s neighborhood relations and spatial locality in their latent higher-level feature representations. In 2015, He et al. adopted the residual block structure and proposed a deep residual network ResNet [2] for image recognition. In 2017, Hu et al. proposed SENet [5] based on channel attention module. In 2020, Zhang et al. proposed ResNeSt [7] based on the Split attention module. In addition, based on the existing CNNs [1]-[7]. The proposal of numerous excellent CNNs makes it possible to classify brain tumor MR images. In general, deep learning is a typical data-driven technology. Especially for meadical tasks, not only the labeling data is expensive, but there is also the problem of scarcity of dataset. For this situation, a favorable substitute for training CNNs from scratch is to use a pre-trained model based on transfer learning [8]. It applies knowledge and skills learned in previous domains or tasks to novel domains or tasks. 

Currently, there are plenty of literatures based on CNNs to achieve classification of brain tumor images [10]. Here we review the most recent and related research works on the paper topic. In 2018, Ge et al. proposed a novel multi-scale CNN architecture to realize the grading of gliomas and got the accuracy of 89.47\%[11]. In 2019, Swati et al. used a pre-trained CNN model and proposed a block-wise ﬁne-tuning strategy to classify glioma, meningioma, and pituitary [12]. It achieved the accuracy of 94.82\%. In 2020, Zhang et al. used the Pyramid Scene Analysis Network to realize the detection and grading evaluation of meningiomas [13]. The accuracy of meningioma detection and three grades classification are 100\% and 81.36\%. Although these methods can perform well in predicting the presence, type, or grade on primary intracranial tumor dataset, they rarely involve secondary intracranial tumor images. Moreover, to the best of our knowledge, there is still a technical gap in distinguishing the source type of tumor (primary or secondary intracranial tumor), which needs to be resolved urgently.

This paper proposes an experimental method for classifying the presence or absence of tumors and the source of tumors. Our scheme is to send tumor images into our proposed network ResNeSAt. Under the prior knowledge provided by transfer learning, the model is fine-tuned to achieve the best performance. Unlike traditional methods that only target a specific type of tumor, our task is more complex and challenging. For our brain MR image dataset, we achieve a higher accuracy than previous methods. In addition, we have also achieved certain results in the new application domain of distinguishing the types of tumor sources.

\section{MATERIAL}

In this paper, we used the new brain MR dataset provided by Ruijin Hospital, Shanghai Jiao Tong University School of Medicine. All experiments were performed in compliance with the Declaration of Helsinki. Written informed consent was acquired from each patient or next of kin. Patients underwent MR scans on a 3T scanner using a 16-channel head coil in a supine position. In order to facilitate the therapeutic planning and treatment of GKRS, a gamma knife rigid head frame (Elekta, Stockholm, Sweden), which was matched with the head coil, was fixed on patient’s head. The T1 MPRAGE sequence was performed on all of the patients 5 minutes after administration of 0.1 mmol/kg body weight of contrast agent (Gadoxetic Acid Disodium Injection, Primovist, Bayer Vital GmbH, Leverkusen, Germany). The dataset contains 3670 T1 MPRAGE sequence brain MRI’s images and includes 7 types of tumors: cavernous hemangioma, glioma, meningioma, acoustic neuroma, colorectal cancer brain metastasis, lung cancer brain metastasis, breast cancer brain metastasis. And it covers 70 cases, patients with brain tumors and patients without brain tumors. The physician provides complete manual annotation for each image. Table I lists the number of images of each class in the dataset. Figure 1 shows a sample of various tumors of the dataset used in this paper.

\renewcommand{\arraystretch}{1} 
\begin{table}[htb]  
  \centering
  \caption{Details of Used Image Dataset.}  
    \scalebox{0.5}{ 
    \begin{tabular}{cccccc}  
    \toprule  
    {\bf Presence of tumor}&{\bf Source type}&{\bf Class}&{\bf Number of patients}&{\bf Number of MR images}\cr  
    \midrule
    Yes&Primary intracranial tumors&Cavernous hemangioma&3&75\cr  
    &&Glioma&5&316\cr  
    &&Meningioma&5&78\cr  
    &&Acoustic neuroma&2&38\cr  
    &Secondary intracranial tumors&Colorectal cancer brain metastasis&7&226\cr  
    &&Lung cancer brain metastasis&29&818\cr
    &&Breast cancer brain metastasis&8&285\cr
    No&-&-&11&1834\cr
    \midrule  
    {\bf Total}&{\bf -}&{\bf -}&{\bf 70}&{\bf 3670}\cr  
    \bottomrule  
    \end{tabular}
    }
\end{table} 

\begin{figure}[htb]
  \centering
  \includegraphics[scale=0.24]{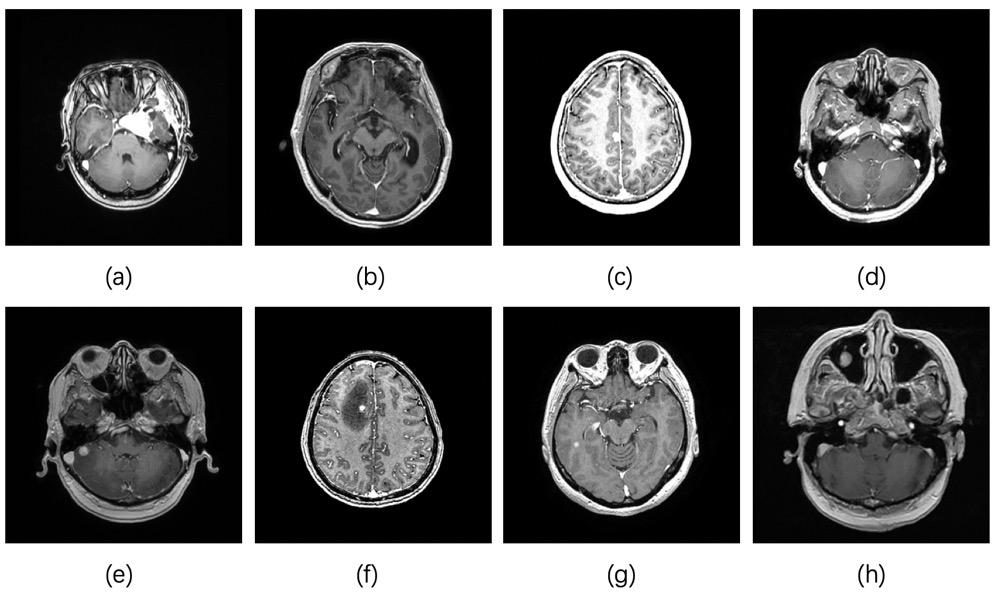}
  \caption{Sample from the used dataset. (a) Cavernous hemangioma, (b) Glioma, (c) Meningioma, (d) Acoustic neuroma, (e) Colorectal cancer brain metastasis, (f) Lung cancer brain metastasis, (g) Breast cancer brain metastasis, (h) No tumor.}
  \label{figurelabel}
\end{figure}

\section{METHODOLOGY}

We propose an experimental method for classifying the presence or absence of tumor and the source of tumor (primary or secondary intracranial tumor) based on brain MR images. The model is trained and optimized through the existing artificially annotated dataset to realize the classification of new images. In this section, the following sub-sections will introduce proposed methodology in detail.

\subsection{Proposed Method} 

\begin{figure}[htb]
  \centering
  \includegraphics[scale=0.14]{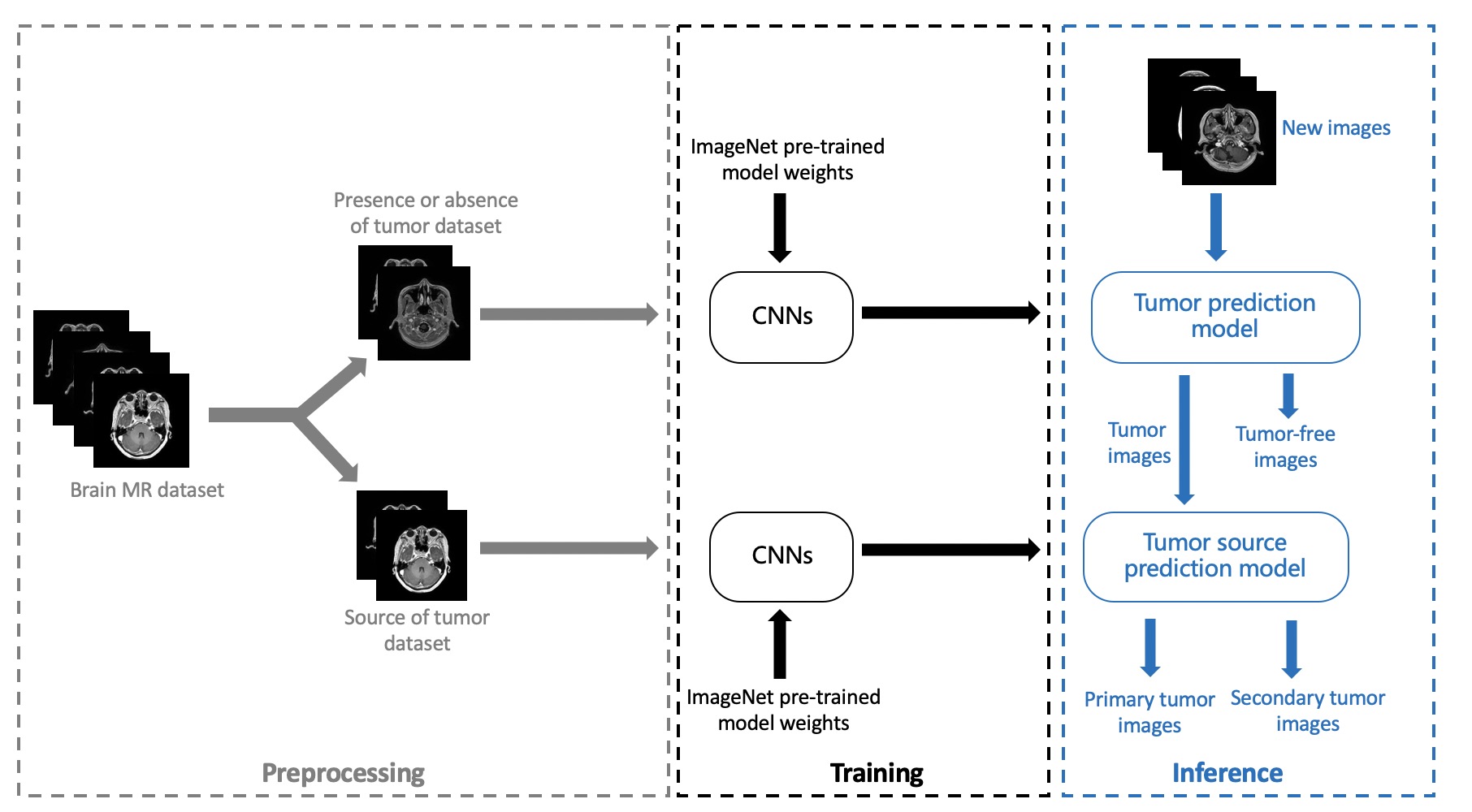}
  \caption{Flow chart of the proposed method.}
  \label{figurelabel}
\end{figure}

Figure 2 shows the flow chart of the proposed method. First, the images in the brain MR dataset are selected into the presence or absence of tumor dataset and the tumor source type dataset according to tumor presence and source type labels. In the preprocessing stage, we use the bilinear interpolation method to unify the image size to 256×256. And we randomly flip 50\% of the pictures horizontally. In this way, although the position of the pixels in the image is changed, the features are kept unchanged, which can enhance the generalization ability of the model. In the following CNN model, the values of pixels in the image will be converted into features. Therefore, the classification effect is dependent on these pixel values.We use normalization to limit the data to a certain range we need. This can prevent the unbalanced image pixel intensity distribution from interfering with subsequent processing. After preprocessing, we send the image to the model for traning. For large-scale network model with huge parameters, when performing medical tasks with relatively rare dataset, there is a certain gap in task performance between using pre-trained model and training from scratch [14]. We used the ImageNet pre-trained model based on pytorch. In each round of iterative training, the model parameters are automatically adjusted based on the prediction results of the previous round. The maximum number of training iterations is set to 100 epochs. We use SGD optimizer to update the weights of the network with a batch size of 16. And an initial learning rate of $\alpha_0$ at the very beginning and cosine decays are as following:
\begin{equation}
\alpha = \frac{1}{2} \times(1+cos(\frac{e\times\pi}{N_e}))\times \alpha0
\end{equation}
where $e$ is an epoch counter, and $N_e$ is a total number of epochs. The $\alpha_0$ for presence or absence of tumor model and source of tumor model are 1e-3 and 1e-4 respectively. After the two models are trained, the two optimized prediction models can be used to sequentially classify the presence or absence of tumors and the type of tumor source on the new image. The model is mounted on Huawei G5500 series servers and uses 1 NVIDIA V100 GPU card. The implementation of our network is based on PyTorch. 

\subsection{Existing Classification Networks Comparison} 

We compare the effects of commonly used CNNs in recent years in the classification of tumor presence and source. 80\% of brain MR dataset were used to train the model, and the remaining images were used to test the model. Finally, the testing accuracy is as shown in Figure 3.

\begin{figure}[htb]
  \centering
  \includegraphics[scale=0.15]{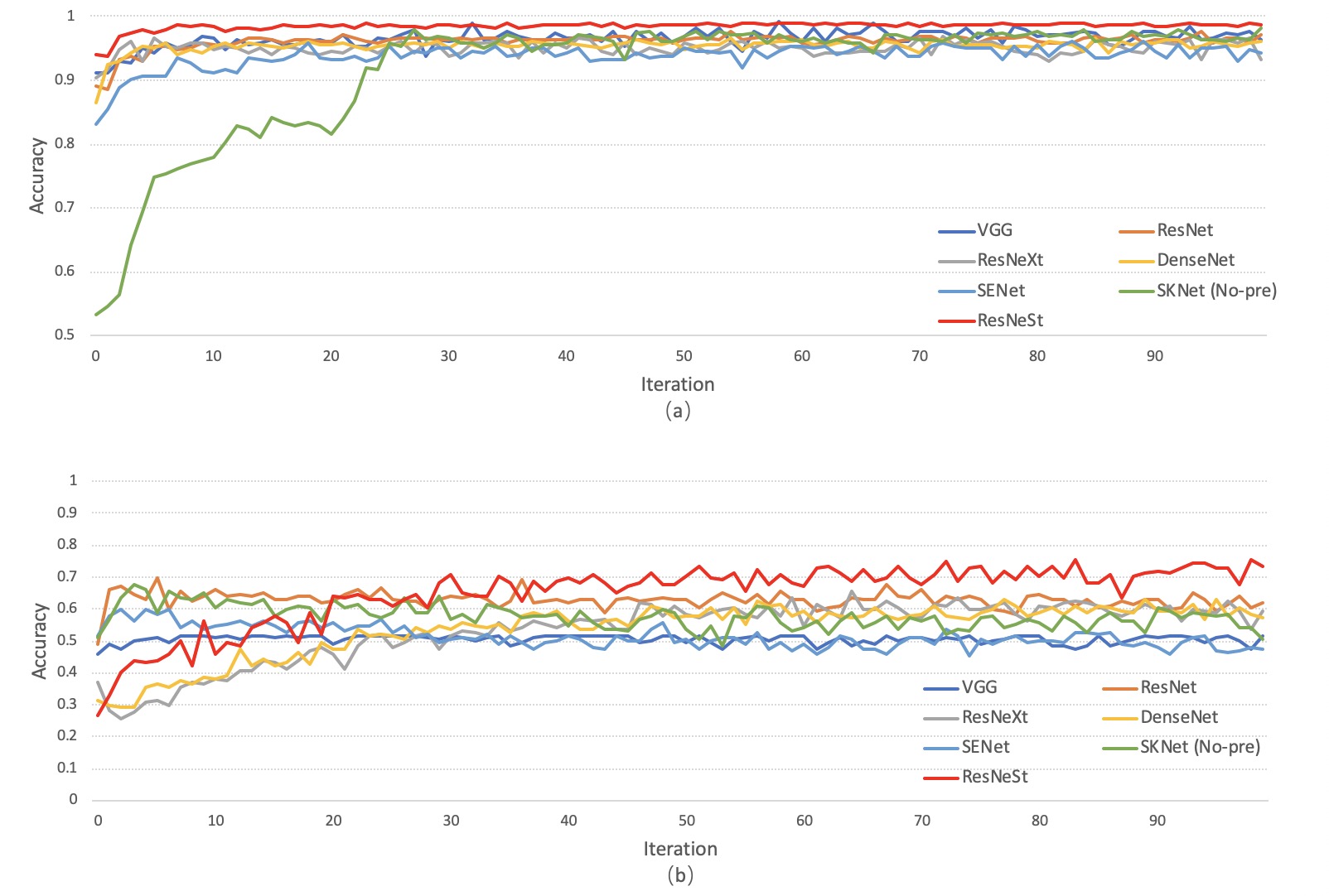}
  \caption{Baseline Comparison. (a) Presence or absence of tumor, (b) Source of tumor.}
  \label{figurelabel}
\end{figure}

It can be seen that on the two classification tasks, the red line performs better. Especially in the classification task of tumor source type, most models cannot effectively distinguish the features between tumors from different sources, but ResNeSt performs well. Therefore, we decided to make the model architecture improvement based on this network to achieve a higher level of task performance.

\subsection{Proposed Architecture ResNeSAt}

Our method is improved on ResNeSt [7]. We borrowed its split mechanism, so that multiple convolution kernel branches in the same layer can extract features separately, making the features more diverse. At the same time, our method uses the channel attention mechanism, which is achieved by assigning different weights to channels of different branches. Next, we use the spatial attention module (SA) in CBAM [9], so we call the proposed model ResNeSAt. SA generates a spatial attention map by concatenating the global pooling and maximum pooling and then convolving to extract the spatial relationship between the elements. In ResNeSAt, we inserted SA into each bottleneck in order to make the entire network pay more attention to the tumor area when extracting features, as shown in Figure 4. Where $\bigotimes$ and $\bigoplus$ represent element-wise multiplication and element-wise summation respectively.

\begin{figure}[htb]
  \centering
  \includegraphics[scale=0.1]{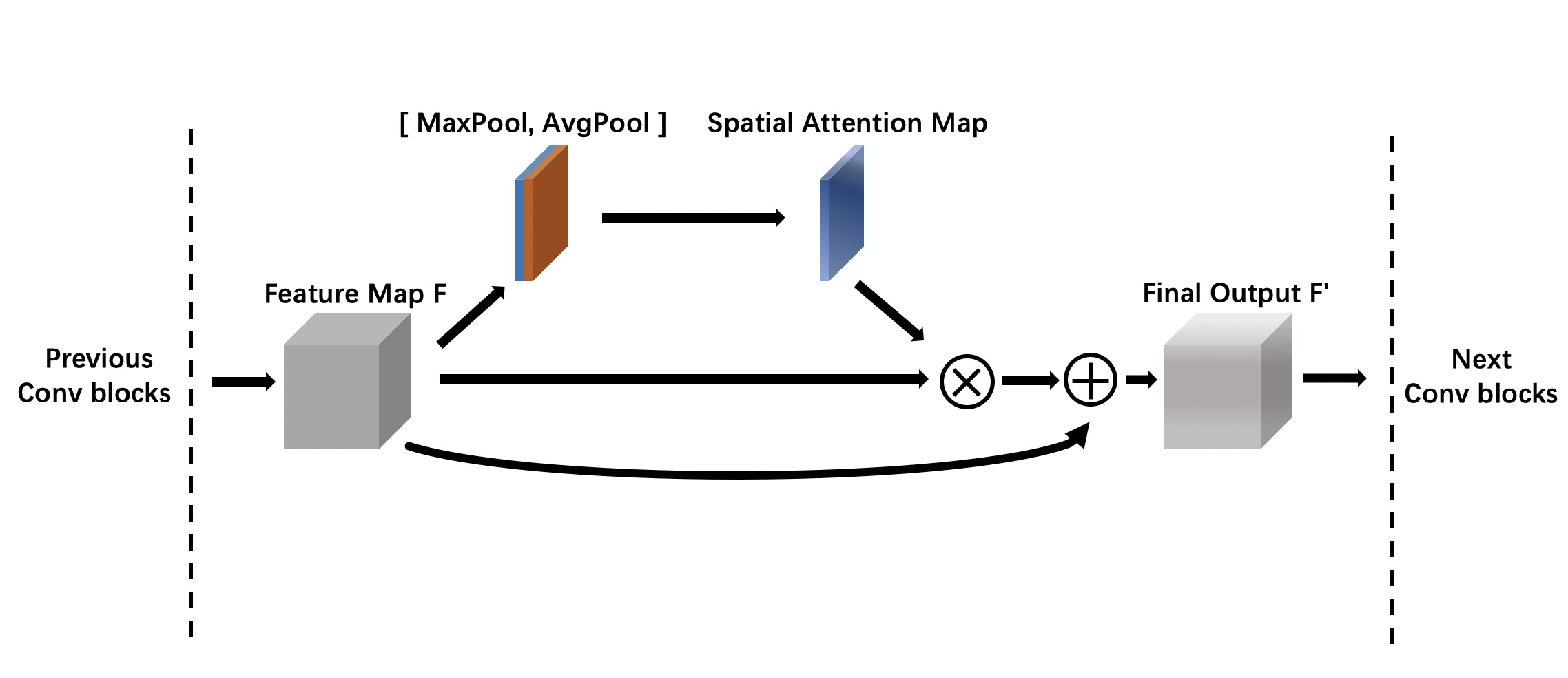}
  \caption{SA integrated with Bottleneck in ResNeSAt.}
  \label{figurelabel}
\end{figure}

\renewcommand{\arraystretch}{1.1} 
\begin{table*}[htb]  
  \centering
  \caption{Details of network ResNeSAt.}  
    \scalebox{1.1}{ 
    \begin{tabular}{cccc}  
    \toprule  
    {\bf ResNeSAt}&{\bf Name}&{\bf Details}&{\bf Repeat}\cr 
    \hline
    &Input&&\cr
    \hline
    InitConv&InitConv 1&Conv3, BN, ReLU&2\cr  
    &InitConv 2&Conv3, BN, ReLU, MaxPool&1\cr  
    \hline  
    Layer 1&Bottleneck 1&Conv1, BN, SplAtConv, Conv1, BN, ReLU, SA, AvgPool, Conv1, BN&1\cr
    &Bottleneck 2&Conv1, BN, SplAtConv, Conv1, BN, ReLU, SA&2\cr  
    \hline
    Layer 2&Bottleneck 3&Conv1, BN, AvgPool, SplAtConv, Conv1, BN, ReLU, SA, AvgPool, Conv1, BN&1\cr  
    &Bottleneck 4&Conv1, BN, SplAtConv, Conv1, BN, ReLU, SA&3\cr  
    \hline
    Layer 3&Bottleneck 5&Conv1, BN, AvgPool, SplAtConv, Conv1, BN, ReLU, SA, AvgPool, Conv1, BN&1\cr
    &Bottleneck 6&Conv1, BN, SplAtConv, Conv1, BN, ReLU, SA&5\cr
    \hline
    Layer 4&Bottleneck 7&Conv1, BN, AvgPool, SplAtConv, Conv1, BN, ReLU, SA, AvgPool, Conv1, BN&1\cr
    &Bottleneck 8&Conv1, BN, SplAtConv, Conv1, BN, ReLU, SA&2\cr
    \hline
    &Classifier&GlobalAvgPool, FC&1\cr
    \bottomrule  
    \end{tabular} 
    } 
\end{table*}  

\subsection{Performance Evaluation}
We have evaluated the classification performance based on recall, specficity, precision, F1-score, and accuracy. To evaluate these metrics, we use four indices: Truly Positive (TP), False Positive (FP), False Negative (FN) and True Negative (TN) and deﬁne the following mathematical expressions.

\begin{equation}
Recall = \frac{TP}{TP + FN} 
\end{equation}
\begin{equation}
Specificity = \frac{TN}{TN + FP} 
\end{equation}
\begin{equation}
Precision = \frac{TP}{TP + FP} 
\end{equation}
\begin{equation}
F1-Score = 2 \times \frac{Recall \times Precision}{Recall + Precision} 
\end{equation}
\begin{equation}
Accuracy = \frac{TP + TN}{TP + TN + FP +FN} 
\end{equation}
 
Recall and Specificity describe how good the classifier is at classifying the positive and negative condition. Precision is the positive predictive rate. F1-score measure the classiﬁcation performance in term of recall and precision. Accuracy is the overall classiﬁcation accuracy of the proposed method.

\section{EXPERIMENTAL RESULTS}

The variability of a single model can be quite high. We evaluated the classiﬁcation performance of the proposed model ResNeSAt and compared it with ResNeSt on ten-fold cross-validation and summarized our results in the form of tables. Table III and Table IV show the average recall, speciﬁcity, precision, F1-score and accuracy of best performed proposed CNN model on ten-fold testing datasets.

As shown in Table III, we can see that our proposed model ResNeSAt has improved recall, specificity,precision, F1-score and accuracy in classifying the presence or absence of tumor. This shows that ResNeSAt pays more attention to the tumor area, so it can better judge whether there is a tumor in the image. As shown in Table IV, in terms of judging the type of tumor source, we found that the improved model ResNeSAt has improved recall, F1-score, and accuracy. Specificity and precision can be almost ignored compared to the more comprehensive indicators F1-score and Accuracy. This shows that when judging the source type of the tumor, ResNeSAt better analyses the details of the tumor area and improves the classification performance to a certain extent. The parameter of the ResNeSt is 97.45 MB, and the parameter of the ResNeSAt is only 97.46 MB. In general, ResNeSAt effectively improves the accuracy of classification of tumor presence or absence and tumor source type, almost without increasing the model size.

\renewcommand{\arraystretch}{1} 
\begin{table}[htb]  
  \centering
  \caption{Presence or absence of tumor model.} 
    \scalebox{0.9}{ 
    \begin{tabular}{cccccc}  
    \toprule  
    &{\bf Recall}&{\bf Specificity}&{\bf Precision}&{\bf F1-score}&{\bf Accuracy}\cr  
    \midrule
    {\bf ResNeSt}&0.9733&0.9772&0.9803&0.9768&0.9751\cr  
    {\bf ResNeSAt}&0.9864&0.9981&0.9983&0.9923&0.9918\cr
    \bottomrule  
    \end{tabular} 
    }
\end{table} 

\renewcommand{\arraystretch}{1} 
\begin{table}[htb]  
  \centering
  \caption{Source of tumor model (primary or secondary).} 
    \scalebox{0.9}{ 
    \begin{tabular}{cccccc}  
    \toprule  
    &{\bf Recall}&{\bf Specificity}&{\bf Precision}&{\bf F1-score}&{\bf Accuracy}\cr  
    \midrule
    {\bf ResNeSt}&0.7927&0.8684&0.8718&0.8304&0.8283\cr  
    {\bf ResNeSAt}&0.8212&0.8480&0.8591&0.8397&0.8338\cr
    \bottomrule  
    \end{tabular} 
    }
\end{table} 

\section{CONCLUSIONS \& FUTURE WORK}

In this paper, we use an attention guided deep learning method to solve the problem of tumor source classification, and further improve the effect of classifying the presence or absence of tumor. Using clinical case data generated by Shanghai Ruijin Hospital, we propose the ResNeSAt network for training and optimization, and classify the presence or absence of tumor and the type of tumor source on the tumor image within the error tolerance. This method produces a fast and accurate differential diagnosis with less subjective bias. It can reduce the dependence on experts in the clinical tumor image prediction process, and improve the efficiency and accuracy of diagnosis. Meanwhile, there is still room for improvement in the classification task of tumor source type. In future work, we can try segmentation of key areas to capture small differences between different images, which may aid to differentiate tumors classification further.

\section*{ACKNOWLEDGMENT}
The author would like to thank each patient or next of kin who provided tumor image data. Also, this research was supported by the high performance computing (HPC) resources at Beihang University.


\end{document}